# Fully-automatic laser welding and micro-sculpting with universal *in situ* inline coherent imaging

Paul J.L. Webster[1,2], Logan G. Wright[1,3], Yang Ji[1], Christopher M. Galbraith[1], Alison W. Kinross[1], Cole Van Vlack[1,2] and James M. Fraser[1*]

1. Department of Physics, Engineering Physics and Astronomy, Queen's University,
Kingston, Ontario K7L 3N6, Canada
2. Laser Depth Dynamics Inc., 2 Gore Street, Kingston, Ontario K7L 2L1, Canada
3. School of Applied and Engineering Physics, Cornell University, Ithaca, New York 14853, USA.
*Corresponding author: fraser@physics.queensu.ca

Though new affordable high power laser technologies make possible many processing applications in science and industry, depth control remains a serious technical challenge. Here we show that inline coherent imaging, with line rates up to 312 kHz and microsecond-duration capture times, is capable of directly measuring laser penetration depth in a process as violent as kW-class keyhole welding. We exploit ICI's high speed, high dynamic range and robustness to interference from other optical sources to achieve fully automatic, adaptive control of laser welding as well as ablation, achieving micron-scale sculpting in vastly different heterogeneous biological materials.

Laser processing offers technical benefits in a wide range of possible manufacturing applications and is emerging as a versatile tool for modern microfabrication, joining, surgery and micromanipulation [1-7]. At a first pass, laser processing can ostensibly be tailored for just about any required modification of any material, in practice limited only by diffraction and the duration of light exposure. However, in long-pulse (pulse duration much greater than 10 ps) laser cutting and laser welding material modification is virtually volcanic, characterized by explosive chemical, thermal and fluid dynamics. Precise modification without deleterious effects requires careful management of light delivery and assist/cover gas flow [2]. With multicomponent materials (e.g., coatings, alloys, microprocessors), or with plasma, gaseous emissions and assist gas interactions, a variety of chemical reactions can arise that completely change the light-matter interaction. By use of ultrafast pulses (pulse duration below about 10 ps [7]), material modification may become more precise with fewer thermal side effects. However, this occurs only within a relatively small parameter space (usually at fluences lower than required for high throughput industrial applications), and is sensitive to any obstruction of light delivery (e.g., newly-machined features, material defects, dispersion, nonlinear distortions, plasma shielding [7-10]). Hence in well-studied and controlled conditions, ultrafast pulses provide exquisite ablation performance, but presently restrict flexibility in geometry and processing speed. These issues cannot be circumvented by improved laser source technology alone.

Accordingly, researchers and manufacturers using high power long pulse light sources are forced to conduct laborious explorations of many-dimensional parameter spaces (e.g., assist gas characteristics, laser wavelength, pulse energy, duration and repetition rate) and to choose conservative operating conditions and high cost feedstock and fixturing in order to achieve reliable accuracy, all without necessarily providing a systematic understanding. Furthermore, these approaches are fundamentally limited by the multitude of inevitably uncontrolled variables present in heterogeneous materials or high-volume manufacturing environments. Many researchers [9-16] have made steps to avoid this lengthy and costly search; they have aspired to observe laser processing *in situ*, both to understand the interactions and to achieve the "so-far elusive goal" [11] of online process control. The challenge has been to find an imaging technique which is as versatile as laser processing and yet still overcomes several unique challenges. Imaging must be done at high enough speed to resolve the fast changing morphology but must have sufficient sensitivity and dynamic range to resolve faint signals from deep, high-aspect ratio features, and be sufficiently robust to reject intense optical process noise.

In 2010, Webster et al. demonstrated a technique called inline coherent imaging (ICI), wherein a spectral-domain low-coherence interferometer was aligned coaxially with a processing laser to observe laser percussion drilling *in situ* [17]. ICI is closely related to spectral-domain optical coherence tomography (SD-OCT) [18]. Surprisingly, the non-contact deployment, high dynamic range and micron resolution that make SD-OCT successful in ophthalmology are equally attractive features for control in the radically different conditions of laser processing. ICI simultaneously resolves backscatter from multiple depths along the laser processing beam path with micron resolution, has high dynamic range (>60 dB), and is robust to all other optical signals (machining light, plasma, blackbody radiation, etc.) due to its inherent spectral filtering and coherent time-gating. ICI and similar techniques have since been applied to manual inspection for industrial [16] and surgical applications [19]. Like biomedical OCT [18], ICI may be implemented cheaply, robustly and compactly using components developed for telecommunications and machine vision applications [20].

In this Letter, we show that by exploiting novel ICI image processing techniques in conjunction with high-speed CMOS technology, ICI can be harnessed to solve a key industrial problem, namely direct measurement of laser weld depth, as well as provide novel laser ablation control to achieve fully-automatic 3D subtractive processing.

Laser keyhole welding is a technique of potential importance for production of photovoltaics [3], electronics [4], as well as automotive and aerospace components [5], and medical devices [6]. However, it lacks a technique to directly measure weld depth *in situ*. In keyhole laser welding, a laser is translated relative to the joint-to-be, creating a (primarily) gas-phase capillary that channels light deep into the workpiece (Figure 1). Efficient and quality welding to any given depth is a delicate balance between the evaporation and recoil pressures (from the absorbed laser power) that create the keyhole and the surface tension driving its collapse. Finding this balance is challenging [21] and sensitive to small changes in parameters. The process is inherently unstable [5,22] and many of the symptoms are only apparent through *ex situ* destructive mechanical testing. Unlike other welding approaches (ultrasonic, resistance), laser welding has lacked a universal tool for rapid, direct, quantifiable process monitoring and automated closed-loop control [14,23]. In spite of the technical advantages of laser welding [2,5], this weakness has been a key impediment to its development in many applications [6]. Accurate weld-depth measurement is required to ensure weld quality, especially with dissimilar/coated materials [5,23] and in microwelding [3].

Here, we show that ICI is a precise tool for real-time, high speed weld depth measurement, even for kW class keyhole welding. A schematic diagram of our experiments is shown in Figure 1. The ICI system is composed of a fiber-optic Michelson interferometer illuminated with a superluminescent diode (SLD, centre wavelength 843 nm, 25 nm FWHM). The fiber-optic beamsplitter splits the SLD light into a free-space reference arm and sample arm, where the imaging light is combined with the processing laser (CW, 1.1 kW max, Yb:fiber, 1070 nm) on a dichroic mirror. Both beams are focused onto the sample using the welding laser objective. The keyhole depth (plotted in yellow) is retrieved from the spectral interferogram resulting from interference between sample and stationary reference arm light. This interferogram is measured on the static spectrometer, which uses a CMOS array sensor for high-speed imaging.

The current gold standard weld depth measurement technique is *ex situ* destructive testing, namely grind, polish, and etch stages followed by brightfield microscopy of weld cross-sections. A systematic comparison of these measurements to *in situ* results from ICI shows excellent agreement (Figure 2). Transverse section measurements were acquired from the depth of the solid weld *ex situ*, as in the inset (measured virgin surface to deepest point, indicated by red dot).

While this implies that ICI accurately measures depth in well-controlled welding, a more difficult task is tracking complicated keyhole dynamics. Figure 3 shows ICI tracking of power-cycled welding (230 kHz imaging rate with 1.5 µs exposure). Starting at 0 s, the power was changed every 25 ms in the sequence 0.96, 3.2, 1.9, 3.2, 0.96 (units $MW/cm^2$) and repeated. The bottom of the welding keyhole (green dots) was tracked by identifying the deepest interface 20 dB above the noise floor within a group of ICI axial images. For comparison, ICI tracked depths are overlaid on the longitudinal cross-section obtained by destructive analysis. Both macroscopic and microscopic details of the weld reveal excellent tracking by ICI, including the sharp jumps in the keyhole's depth caused by changes in the laser power but also the unpredictable variations apparent on a smaller scale. These results show how ICI may be useful for real-time identification of pathological behavior in a keyhole welding process (here, porosity), which is essential for online quality control.

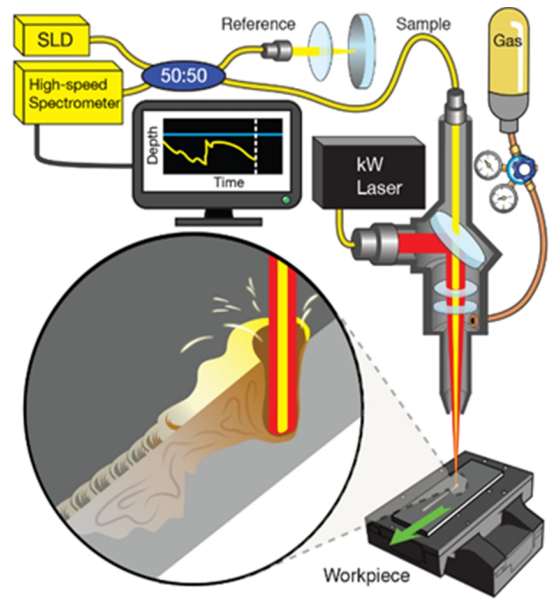

Fig 1: Schematic showing ICI system integrated into a laser welding work station. SLD – superluminescent diode (yellow beam). 50:50 – fiber beamsplitter. Green arrow indicates material feed direction.

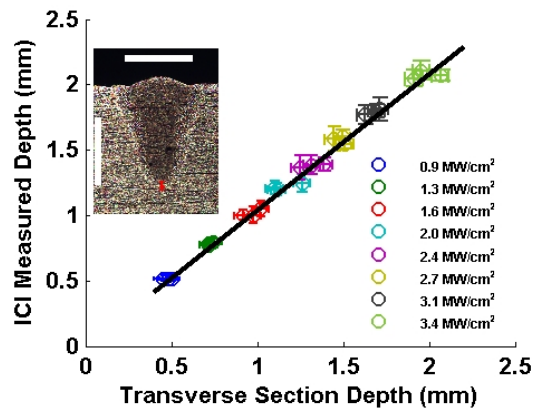

Fig 2: *In situ* depth measurements compared to destructive analysis for bead-on-plate welding (mild steel, spot size 210 µm, material feed rate 40 mm/s, argon cover gas) with linear fit (slope 1.04±0.04). (inset) Brightfield microscope image of a weld cross-section (for 2.7 $MW/cm^2$)). Inset scale bars correspond to 1 mm.

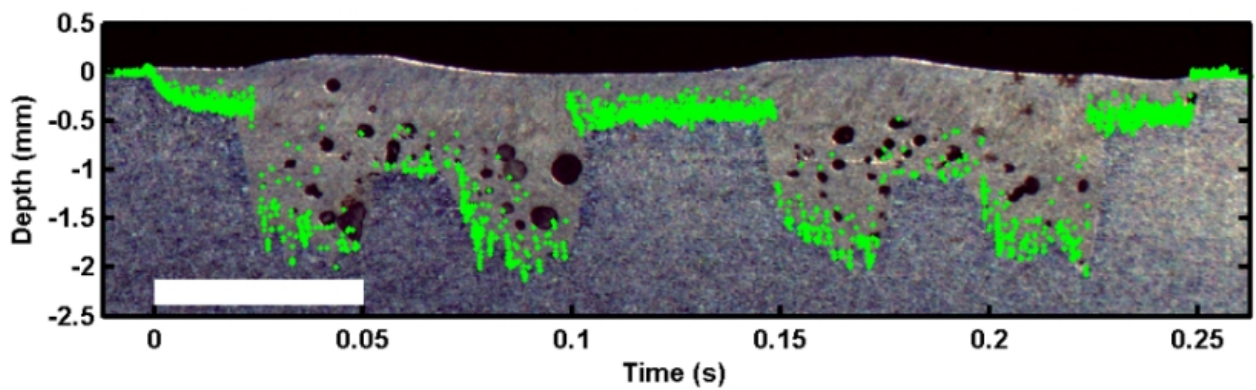

Fig 3: ICI measured depths (green dots) due to power-cycled weld (details in text), overlaid on brightfield microscopy image of longitudinal cross-section. Scale bar corresponds to 2 mm.

*In situ* monitoring of laser processing is possible with a variety of techniques, but automatic laser processing (i.e., closed-loop control) has more stringent technical requirements. Measurement must be robust to high speed instabilities (e.g., temporary constriction of the keyhole vapour capillary) and backscatter from many depths, without requiring extensive on-the-fly data processing so as to provide timely feedback. An additional challenge arises for imaging that involves a spatially coherent beam; backscattered intensity varies strongly due to speckle arising from submicron spatial variations in the sample morphology. High speed changes in the morphology can also defeat ICI and other interferometric-based imaging techniques since contrast is lost for interfaces moving at speeds approaching the wavelength of the imaging light per integration time (so called "fringe wash-out" [24]). We overcome these challenges by collecting bursts of short integration time images (to minimize fringe wash-out) that are processed to obtain one depth measurement per group (to reduce speckle and processing time requirements). This measured depth is compared to the target depth. Integral gain feedback based on the depth mismatch yields fast response to changes in the desired target depth, or alternatively changes in the work piece composition (Figure 4). Without feedback, constant power and scan rate leave a gap between the top layer and substrate (Figure 4a) resulting in a failed lap weld. Closed-loop control using ICI feedback (controlling the laser power) quickly achieves the target depth, achieving fusion with no gap (Figure 4b) with a slight overshoot that is self-corrected. For this particular implementation, the response time of the control signal is 18 ms (1/e) which is sufficient to remove the gap between plates. The gain factor of the integral control loop was set to maximise response speed while minimizing overshoot.

Laser additive manufacturing has generated considerable excitement for its ability to create components with user-defined 3D morphology through a direct write process. It is, however, severely limited by the final material characteristics which are strongly influenced by the laser fusion or polymerization process. Using ICI integrated into a laser microprocessing system, we achieve three-dimensional, fully-automated milling of materials. Here, the material properties of the final component are essentially the materials' virgin properties. We demonstrate this approach with contrasting heterogeneous materials to illustrate its flexibility.

Our experimental setup is similar to Figure 1, except the welding laser is replaced by a 6 ps, 355 nm, 50 µJ diode-pumped solid state laser focused to an 18 µm $e^{-2}$ intensity diameter. First, a target shape is designed (Figure 5 top). During sample XY scanning, *in situ* depth measurements are compared to the target shape and automated feedback adaptively controls the number of laser pulses to each point, to generate the final shape in bovine cortical bone (Figure 5 left). When the process is repeated with a different material (wood, Figure 5 right), the necessity of closed-loop control is made especially clear. A narrow band of latewood (L in Figure 5 right, dense cell structure) had a mean etch rate of 4.0±0.4 µm/pulse, lower than that in the large-cell-structured, nearby earlywood (E, 14±5 µm/pulse). These variations were automatically compensated and quantified by ICI control. Both sculptures shown were produced on the first attempt in each material, illustrating the potential to virtually eliminate specific process development.

We have demonstrated high speed laser keyhole depth control and full, three-dimensional automatic sculpting of highly heterogeneous materials using a relatively simple approach which adapts high-speed coherent imaging to coaxial-observation of laser processing. We have shown how the technique can eliminate the need for—and even vastly exceed the accuracy of—*a priori* mean process parameter optimization techniques. With growing excitement about versatile 3D manufacturing, the uncompromising flexibility, material diversity and simplicity of ICI control make it a prime candidate to enable future laser processing applications.

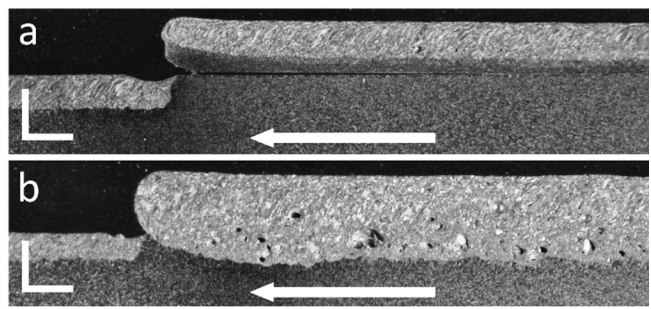

Fig 4: Laser keyhole lap welding of two different mild low-carbon steel plates (white arrow indicates material feed direction) a, Constant power leaves gap between top and bottom plate indicating failed joining (320 W, spot size 210 µm, material feed rate 20 mm/s, argon cover gas). b, Successful weld through closed-loop control with laser power set by ICI feedback. All other parameters match open-loop settings. Scale bars 0.5 mm.

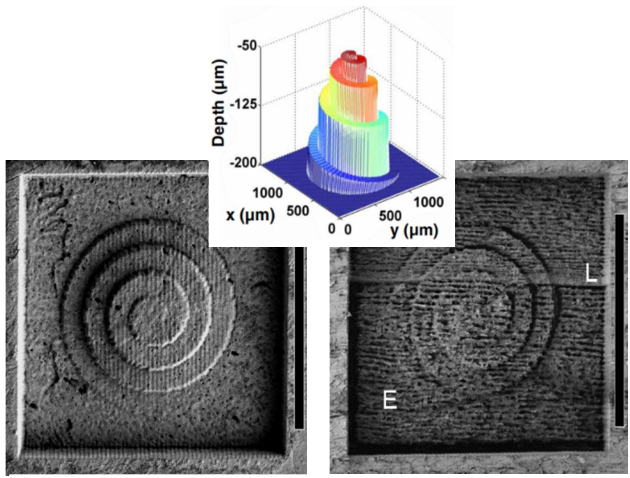

Fig 5: Fully-automatic 3D laser micromachining of heterogeneous biological materials. Top: User-designed target depth map, left: SEM image of bone spiral, right: SEM image of wood spiral showing underlying cellular structure. Wood grain is evident with dense latewood (L) and large-cell earlywood (E). Black scale bars are 1 mm

We thank Kevin Mortimer and Karen Yu for early contributions and helpful discussions, and Charles Cooney for assistance in destructive analysis and SEM imaging. Funding for this work was provided by the Natural Sciences and Engineering Research Council of Canada, the Ontario Centres of Excellence, and the Canada Foundation for Innovation. P.J.L.W and L.G.W conceived the initial ideas for this paper. P.J.L.W. designed and built the inline coherent imaging systems, and with assistance from A.W.K and C.M.G, built the welding apparatus. P.J.L.W., L.G.W., Y.J., and C.M.G. designed and built the ultrafast micromachining station. Welding depth measurement experiments were performed by A.W.K., with code from C.V. and P.J.L.W., data analysis by C.V. and *ex situ* analysis by A.W.K. Welding depth control and analysis was achieved by C.M.G., using code developed by C.V. 3D sculpting was performed by L.G.W., Y.J., C.M.G., and C.V. with code developed and implemented by P.J.L.W., C.V., and Y.J. SEM analysis was completed by C.M.G.. L.G.W. led the preparation of the manuscript, which was discussed and contributed to by all authors. J.M.F. led the research team.